\documentclass{article}
\usepackage[english]{babel}
\usepackage[ansinew]{inputenc}

\usepackage{babelbib}
\usepackage{hyperref}
\usepackage{url}

 \usepackage{bibentry}

\usepackage{graphics, graphicx}
\usepackage{latexsym}
\usepackage{colortbl}
\usepackage[margin=11pt,font=small,labelfont=bf]{caption}

\usepackage{tgadventor}

\usepackage{pdflscape}
\usepackage{pdfpages}
\usepackage{fancyhdr}
\usepackage{mathtools}
\usepackage{multirow}
\usepackage{longtable}
\usepackage{array}
\usepackage{amsmath}
\usepackage{subfigure}
\begin{document}

\title{\textbf{Synthetic Embedding of Hidden Information in Industrial Control System Network Protocols for Evaluation of Steganographic Malware} \\[0.2cm] {Technical Report}}

\author{
  Tom Neubert, Bjarne Peuker, Laura Buxhoidt,\\ 
  Eric Schueler, Claus Vielhauer\\ \\
  Brandenburg University of Applied Sciences\\
  Department of Computer Science \& Media\\
  Brandenburg, Germany \\
  \texttt{firstname.lastname@th-brandenburg.de}
}

\maketitle

\section*{Abstract}
For the last several years, the embedding of hidden information by steganographic techniques in network communications is increasingly used by attackers in order to obscure data infiltration, exfiltration or command and control in IT (information technology) and OT (operational technology) systems. Especially industrial control systems (ICS) and critical infrastructures  have increased protection requirements. Currently, network defense mechanisms are unfortunately quite ineffective against novel attacks based on network steganography. Thus, on the one hand huge amounts of network data with steganographic embedding is required to train, evaluate and improve defense mechanisms. On the other hand, the real-time embedding of hidden information in productive ICS networks is crucial due to safety violations. Additionally it is time consuming because it needs special laboratory setup. \newline To address this challenge, this work introduces an embedding concept to generate synthetic steganographic network data to automatically produce significant amounts of data for training and evaluation of defense mechanisms. The concept enables the possibility to manipulate a network packet wherever required and outperforms the state-of-the-art in terms of embedding pace significantly.   

\clearpage
\section{Introduction}

Nowadays, Industrial Control Systems (ICS) play an essential role in our economy and in our every day life. In various cyber-physical systems they help to control and automate processes. The field of supported automation processes is significant from rather simple traffic light controls to complex networks of ICS (e.g. in nuclear power plants). Since the well known Stuxnet-Attack \cite{Stux13}, the IT-security community is aware that it is of great importance to ensure the further development of security mechanisms of ICS against cyber attackers. In particular with the rise of Ethernet connection and TCP/IP based protocols in ICS, the boundaries between OT (operational technology) and IT (information technology) are blurred. Due to these blurred boundaries, similar threats and attacking trends in the IT-domain can be observed in the OT-domain (e.g. Stuxnet-Attack \cite{Stux13}, Ukrainian Power Grid Attack \cite{UkrainianPG} and attacks on safety systems \cite{schneiderattack}). \newline
Thanks to the continuous research and development of defense mechanisms in ICS, there is a rather reliable and precise functionality against conventional attack vectors. Thus, it is a current trend among attackers to use steganographic techniques to hide information in network communication \cite{MITRE:Stego} to conceal relevant attack procedures (especially) in advanced persistent threats (APTs), like data in- and exfiltration, command \& control communication or lateral movement for as long as possible. Due to the mentioned blurred boundaries between OT and IT, attackers could use hidden channel network communication in ICS as shown in \cite{IFAC2020,ICONS2020,AMNT2020}. For these purposes, current defense mechanisms are unfortunately inadequate to counter such novel attack vectors with sophisticated hidden channel network communication and urgently need further development. \newline For the elaboration and evaluation of novel defense mechanisms, diverse and heterogeneous data for training and test purposes is essential. However, the generation of network data with hidden information based on steganographic techniques in ICS is challenging, because each steganographic embedding needs a complex, sophisticated and very specific experimental setup and implementation which is very time consuming, plus raises various security and safety issues. \newline Due to this, in \cite{ARES21} an approach is presented which enables the possibility to generate network data with steganographic embedding artificially, based on network data recordings. The approach from \cite{ARES21} (see Section 2.2) is limited to a very specific steganographic embedding technique based on TCP-timestamps and additionally it has a proportionate low embedding pace which limits the amount of synthetic data, that can be produced in a given amount of time.  \newline Thus, \textbf{this work introduces} an advanced concept for the synthetic embedding of hidden information in network data, which offers the possibility to embed hidden information literally everywhere in (uncompromized) network packet recordings with a significantly faster embedding pace than the state-of-the-art approach from \cite{ARES21}. From now on, we refer to it as \textbf{S}ynthetic \textbf{S}teganographic \textbf{E}mbedding (SSE) concept. The SSE-concept has two synthetic embedding options, one option focuses on a high embedding pace and the other option on a more comfortable and easier to handle embedding.  \section{State-of-the-Art}
In this Section, we introduce the term \textit{network steganography} in the context of this technical report (Section \ref{sec:network_stego}) and present related work regarding the generation of synthetic network data in Section \ref{sec:asnd}.

\subsection{Network Steganography in ICS} \label{sec:network_stego}
According to \cite{Wendzel_2022} with a focus on network steganography the term \textit{steganography} can be described as follows: ``Steganography is the art and science of concealing the existence of information transfer and storage". Thus, network steganography targets the embedding of hidden data in network communication i.e. traffic. From an attacker's perspective, a warden (e.g. intrusion detection system) observes the network traffic and the embedding should be inconspicuous in a sense that he would not be able to differ between genuine and steganographic communication \cite{IHMMSEC2020}. This can be realized for example by manipulating the packets payload on least significant bits, by using unused space in headers or by artificially produced timing delays to modulate time intervals between specific packets \cite{Mazurczyk2018}. \newline Network steganography in the ICS network communication domain is specific due to the lower amount of available data for potential embedding, because transmitted network packets are usually smaller since only meta-data or few (sensor) values are transmitted to keep the communication lean and simple. Additionally, ICS specific protocols like OPC-UA\footnote{\url{https://opcfoundation.org/about/opc- technologies/opc- ua/}} or Modbus-TCP\footnote{\url{https://www.modbus.org/docs/Modbus_Messaging_Implementation_Guide_V1_0b.pdf}} are very commonly encapsulated in actual transport protocols such as TCP/IP, which enables the opportunity for utilizing the data fields of the ICS specific protocols in addition to TCP/IP protocol headers. It is also not uncommon for the ICS-specific payload to be transmitted unencrypted. \newline Relevant and exemplary attack vectors with steganographic embedding techniques in ICS are presented in \cite{IHMMSEC2020, IFAC2020, IHMMSEC21, ICONS2020}. \newline Potential network steganographic embedding patterns and a related terminology is summarized in \cite{ARES21_W}. A generic taxonomy with the intention of a unified understanding of terms and their applicability for steganographic methods can be found in detail in \cite{Wendzel_2022}.

\subsection{Synthetic Steganographic Network Data Generation} \label{sec:asnd}
In 2021 an initial artificial steganographic network data (ASND) generation concept was introduced in \cite{ARES21} due to the lack of available training and test data to elaborate and evaluate defense mechanisms in ICS against steganographic attack vectors. To counter this research gap, the ASND generation concept promises an easy and fast generation of artificial network data, which simulates a steganographic embedding in TCP-timestamps. It is stated that most important factors to be simulated would be: \begin{enumerate}
	\item the physical network including layout and components,
	\item the network traffic including types of flows, directions, protocols used, typical payloads, etc. and
	\item the type and characteristics of the steganographic hidden channel.
	\end{enumerate}

In \cite{ARES21} and in this work only the last aspect (3) of this list is simulated, the rest is taken directly from a uncompromized recording of a physical laboratory setup, which ensures a plausible environment. For the network data capturing the work uses \textit{Wireshark}\footnote{\url{ https://www.wireshark.org}}. For the steganographic embedding and the manipulation of the network packet captures, a processing pipeline with tools from the \textit{Wireshark} family (editcap, mergecap tshark) is used. \newline However, the original ASND generation concept is limited to a very specific embedding only for TCP-timestamps and has quite low embedding pace with 0.86 seconds per packet on commodity hardware. This paper aims to overcome these shortcomings with a concept that enables the opportunity to embed hidden information literally everywhere in a network packet (plausible or not) with significantly higher embedding pace (instant manipulation of thousands of packets).

\section{SYNTHETIC STEGANOGRAPHIC EMBEDDING (SSE)-CONCEPT} \label{sec:concept}
\begin{figure*}[htb]
 \centering
  \includegraphics[width=1\linewidth]{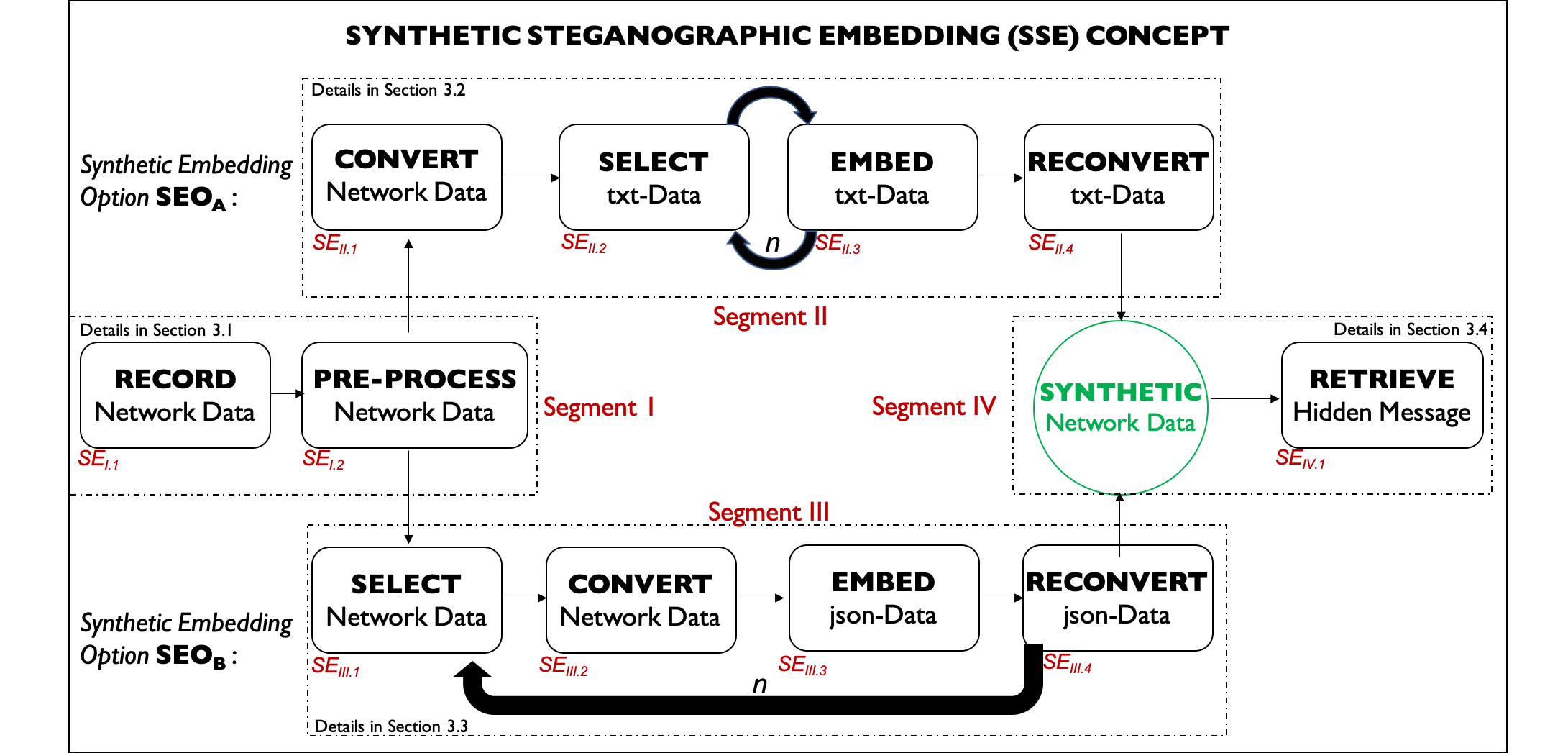}
 \caption{Novel Synthetic Steganographic Embedding Concept for Network Data in ICS}
  \label{img:concept}
\end{figure*}

Our novel Synthetic Steganographic Embedding (SSE)-Concept for network data is based on the idea to enable the opportunity to create appropriate synthetic steganographic network data in an straightforward, comprehensible, fast and easy way for the training and evaluation of defense mechanisms against modern attack vectors which use steganographic techniques to hide information within the network traffic to stay undetected (comparable to \cite{ARES21}). The novel and advanced SSE-Concept shall offer a significantly faster  and unrestricted embedding in network data than the state-of-the-art. \newline Basically, the concept does not synthesize the complete network traffic, it takes recorded network traffic from an uncompromized ICS (laboratory) setup and selects packets and parts of a packet's payload to synthesize only a chosen and specific part of the payload which is well suited for a steganographic embedding of hidden information (e.g. sensor values or timestamps). The rest of the network traffic stays untouched with complete integrity, this ensures a plausible environment so that potential defense mechanisms can focus on the impact of the used steganographic embedding (as mentioned in Section \ref{sec:asnd}). The SSE-concept is visualized in Figure \ref{img:concept} and has two synthetic embedding options (SEO$_A$ and SEO$_B$). SEO$_A$ focuses on a very fast and efficient embedding without accessing structural elements of a packet, the processing is based on parsing through the \textit{hexdump} of packets. The other embedding option SEO$_B$ delivers a much more comfortable embedding with an easier access to structural elements of a network packet based on \textit{json}-objects, but the process is more complex and thus more time consuming. \newline The concept and its four segments: 
\begin{itemize}
	\item Segment I: Record and Pre-Processing
	\item Segment II: SEO$_A$
	\item Segment III: SEO$_B$ and 
	\item Segment IV: Retrieval
\end{itemize} are described in detail in the following subsections. 
\subsection{Segment I: Record and Pre-Process Network Data}
The concept starts with Segment I and its first segment element $SE_{I.1}$ where initially network data is captured from an uncompromized laboratory ground truth (ICS-) setup. The network data can be recorded with different tools, we use \textit{Wireshark} for the recording. The output should be delivered in \textit{pcap} or \textit{pcapng} format\footnote{\url{https://www.tcpdump.org/}} for further processing. These file formats are common and widespread logging protocols which are well suited for the structural recording of network data over a specific time period. \newline In the following   step $SE_{I.2}$ a pre-processing is performed in order to identify relevant communication between selected partners of a specific communication where a steganographic embedding shall be performed. For $SE_{I.2}$, we use \textit{Wireshark} filters.  After the recording and pre-processing of the network data the resulting \textit{pcap}- or \textit{pcapng}-file with the packet capturing can now be used for synthetic embedding and forms our steganographic cover. Based on the created cover, the embedding can be performed with the synthetic embedding options SEO$_A$ (see Section 3.2) or SEO$_B$ (see Section 3.3).

\subsection{Segment II: Synthetic Embedding Option (SEO) A}
Segment II contains SEO$_A$ (\textbf{S}ynthetic \textbf{E}mbedding \textbf{O}ption) where the network traffic can be artificially manipulated based on \textit{txt}-hexdump files. The segment has four elements ($SE_{II.1}$ - $SE_{II.4}$). The first element $SE_{II.1}$ converts the captured network recording to hexdump and saves the hexdump to \textit{txt}-file format. For the conversion, the \textit{Wireshark} console application \textit{tshark} can be used with the following command: 

\begin{verbatim}
tshark --hexdump frames -P -t ad -o 
gui.column.format:
"Time","%t" -r filepath.pcap > output.txt
\end{verbatim}

The command has the following meaning: the term \verb|--hexdump| \verb|frames| means that only all packets bytes are converted to hexdump without secondary data sources like decrypted data to reduce the errors when reconverting the manipulated hexdump back into \textit{pcap}-format. \newline The term \verb|-P -t ad -o gui.column.format:| \verb|"Time","%t"| defines the timestamp format (if not set, packets will have incorrect timestamps) and \verb|-r| is the read option to import a source file. After the conversion into \textit{txt}-data format, relevant packets have to be selected ($SE_{II.2}$) for embedding ($SE_{II.3}$) and therefore, a script for automation has to be elaborated. \newline In detail, by analyzing the previously recorded and pre-processed network traffic, suitable packets for a potential embedding have been selected ($SE_{II.2}$) now. Suitable packets contain payload like sensor values, timestamps or other numeric values with at best many rather irrelevant digits which make an embedding as unobtrusive as possible. The packet selection ($SE_{II.2}$) of suitable packets for potential embedding can be done by identifying unique features (binary signatures) of these packets and finding the position of these packets with the resulting hexdump signature in the \textit{pcap}-file. \textit{Wireshark} offers the possibility to interpret structural packet elements of the encapsulated ICS protocol (like OPC-UA, Modbus-TCP) to interpret the \textit{pcap(ng)}-files. We use this functionality to look up related hexdump signature sequences in order to identify and select suitable packets for potential embedding (\verb|embedding_| \verb|packet_feature|). \newline In the same manner, we identify the actual position of the structural element for embedding ($SE_{II.3}$) of the payload (\verb|embedding_value|) within the previously selected packets. \newline The steps $SE_{II.2}$ and $SE_{II.3}$ are performed in a loop with the elaborated automation script which selects the packets and manipulates the hexdump \textit{txt}-file of the packets until the embedding of the hidden message is complete. This results in a manipulated \textit{txt}-(hexdump) file. In very simplified pseudo code structure these steps could be described as follows: \\
\begin{verbatim}
for hexdump_line in txt_file:
    if hexdump_line.contains(embedding_packet_feature):
        hexdump_line.replace(value, embedding_value)
        
\end{verbatim}
After the completed embedding, the resulting \textit{txt}-file has to be reconverted ($SE_{II.4}$) into a plausible, unconspicous and synthetic network data file. This can be done with the \textit{Wireshark} console application \textit{text2pcap} with the following command: \\
\begin{verbatim}
text2pcap -F pcap -N Eth1 -t "%F %T,%f"  in.txt  out.pcap

\end{verbatim}

where \verb|-F pcap| sets the output file type and \verb|-N Eth1| defines the synthetic interface name of the \textit{.pcap(ng)}-file. \verb|-t "%F %T,%f"| defines the date/time coding, it has to be consistent with the coding from segment element $SE_{II.1}$. The result of Segment II (SEO$_A$) is synthetic network data with hidden information embedded by steganographic techniques.

\subsection{Section III: Synthetic Embedding Option (SEO) B}

Segment III includes SEO$_B$ which offers the possibility to manipulate network packets based on \textit{json}-data. SEO$_B$ is comparable to the embedding process of \cite{ARES21} but instead of the \textit{Wireshark} tool \textit{editcap} it uses \textit{json} file format to manipulate network packets, because \textit{editcap} is limited to TCP-timestamps (as mentioned). Here, steps from $SE_{III.1}$ to  $SE_{III.4}$ are carried out in a loop (in contrast to SEO$_A$) until the embedding of the hidden message is complete. Therefore, a script for automation for these steps has to be elaborated (we use \textit{Python}). Before the start of SEO$_B$, a list with identifiers (\verb|packet_ID|) of suitable packets has to be generated. Suitable packets here have the same characteristics as in SEO$_A$ and can be chosen with \textit{tshark} filters. The script needs access to this created list. SEO$_B$ starts with selecting a single suitable packet ($SE_{III.1}$) for embedding by isolating it with: 

\begin{verbatim}
editcap -r input.pcap isolated_packet.pcap 
            packet_ID

\end{verbatim}

where the term \verb|-r| is the read option followed by the output path and  \verb|packet_ID| is the index number of a selected suitable packet for embedding. Now the isolated packet has to be removed from the initial input file to replace it later with the synthetic packet with: \\
\begin{verbatim}
editcap -V input.pcap input.pcap packet_ID

\end{verbatim}

After the selection and the removal of the isolated packet from the input, the single packet is converted into json file format ($SE_{III.2}$) with: \\

\begin{verbatim}
tshark -r isolated_packet.pcap -T jsonraw >> packet.json

\end{verbatim}

because converting the complete network traffic into \textit{json}-data creates a file that is difficult to process due to a potentially huge file size. Now, a single network packet is represented as a\textit{json}-object and thus it can be processed and manipulated with self-written source code anywhere. The term \verb|-r| defines the read option, the term -T defines the format of the text output. \newline As in SEO$_A$, we recommend to select suitable packet payload for embedding with for example numeric values with many heterogeneous digits which make an embedding as unobtrusive as possible ($SE_{III.3}$). After embedding the information into a single packet \textit{json}-data, the \textit{json}-packet with synthetic embedding is now reconverted with the external \textit{Wireshark} console application \textit{json2pcap} and merged back into the input with \textit{mergecap}, therefore the following commands can be used: \\
\begin{verbatim}
json2pcap -i packet.json -o stego_packet.pcap
mergecap -w output.pcap input.pcap stego_packet.pcap 

\end{verbatim} 
where \verb|json2pcap| reconverts the manipulated packet back into \textit{pcap} network data  and \verb|mergecap| merges it back into the initial input network data cover ($SE_{III.4}$).

\subsection{Segment IV: Retrieval}
In the last Segment IV, the embedded message has to be retrieved from the synthetic network data in order to find out if the embedding was successful ($SE_{IV.1}$). To do so, a retrieval script has to be elaborated with access to the hidden message based on the secret information (from the embedding process like steganographic embedding key) that are necessary for retrieval. Additionally, the generated synthetic network data with steganographic embedding has to be structurally and syntactically analyzed with \textit{Wireshark} to detect and mark obvious structural error in the network data. If the analysis results in no error detections, the synthetic steganographic embedding is successful and completed. \\In the next Sections (4 \& 5) the SSE-Concept is validated by an exemplary embedding and an analysis.

\section{Conclusion and Future Work}
In this technical report we present a novel synthetic steganograhic embedding (SSE) concept with two synthetic embedding options (SEO$_A$ is fast, SEO$_B$ is more flexible in terms of accessing dedicated elements of the encapsulated protocol, both options are equally precise) to generate substantial amounts of steganographic network data in ICS to train and evaluate defense mechanisms in the future work. The novel SSE-concept clearly outperforms the state-of-the-art approach from \cite{ARES21} in terms of embedding pace and delivers the novel opportunity to manipulate a network packet anywhere. The SSE-concept  forms the foundation for our future work, to analyze steganographic data in ICS networks to elaborate suitable defense mechanisms. 

\section*{Acknowledgements}
The presented work is funded by the German Federal Ministry for the Environment, Nature Conservation, Nuclear Safety and Consumer Protection (BMUV, project no. 1501666B) in the framework of the German reactor safety research program.
\bibliographystyle{plain}

\begin{thebibliography}{10}

\bibitem{ICONS2020}
M.~Hildebrandt, R.~Altschaffel, K.~Lamshoeft, M.~Lange, M.~Szemkus, T.~Neubert,
  C.~Vielhauer, Y.~Ding, and J.~Dittmann.
\newblock Threat analysis of steganographic and covert communication in nuclear
  i\&c systems.
\newblock {\em In Proceedings of IAEA ICONS 2020: International Conference on
  Nuclear Security: Sustaining and Strengthening Efforts, 10-14 February 2020,
  Vienna, Austria, \url{https://event.do/iaea/a/\#/events/3301/f/29007}}, 2020.

\bibitem{IHMMSEC2020}
M.~Hildebrandt, K.~Lamshoeft, J.~Dittmann, T.~Neubert, and C.~Vielhauer.
\newblock Information hiding in industrial control systems: An opc ua based
  supply chain attack and its detection.
\newblock pages 115--120, 2020.

\bibitem{schneiderattack}
Dragos Inc.
\newblock Trisis malware - analysis of safety system targeted malware.
\newblock {\em \url{https://dragos.com/wp-content/uploads/TRISIS-01.pdf}},
  2018.

\bibitem{Stux13}
D.~Kushner.
\newblock {\em The Real Story of Stuxnet}.
\newblock IEEE Spectrum; DOI: 10.1109/MSPEC.2013.6471059, 2013.

\bibitem{IHMMSEC21}
K.~Lamshoeft, C.~Kraetzer, J.~Dittmann, T.~Neubert, and C.~Vielhauer.
\newblock Information hiding in cyber physical systems: Challenges for
  embedding, retrieval and detection using sensor data of the swat dataset.
\newblock {\em In Proceedings of the 2021 ACM Workshop on Information Hiding
  and Multimedia Security (IHMMSec '21), June 22--25, 2021, Virtual Event,
  Belgium. ACM, New York, NY, USA, 12 pages.
  https://doi.org/10.1145/3437880.3460413}, 2021.

\bibitem{AMNT2020}
K.~Lamshoeft, T.~Neubert, M.~Lange, R.~Altschaffel, M.~Hildebrandt, Y.~Ding,
  C.~Vielhauer, and J.~Dittmann.
\newblock Novel challenges for anomaly detection in i\&c networks: Strategic
  preparation for the advent of information hiding based attacks.
\newblock {\em Atw. Atomwirtschaft}, 65:504--508, 10 2020.

\bibitem{UkrainianPG}
R.~M. Lee, M.~J. Assante, and T.~Conway.
\newblock Analysis of the cyber attack on the ukrainian power grid.
\newblock Technical report, SANS Institute, 2016.

\bibitem{Mazurczyk2018}
W.~Mazurczyk, S.~Wendzel, and K.~Cabaj.
\newblock {\em Towards Deriving Insights into Data Hiding Methods Using
  Pattern-based Approach.}
\newblock ARES 2018, 13th International Conference on Availability, Reliability
  and Security; Hamburg, Germany, August 27 - August 30, ISBN:
  978-1-4503-6448-5, 2018.

\bibitem{MITRE:Stego}
MITRE and ATTACK.
\newblock Data obfuscation: Steganography.
\newblock {\em
  \url{https://attack.mitre.org/versions/v14/techniques/T1001/002/}}, 2020.

\bibitem{ARES21}
T.~Neubert, C.~Kraetzer, and C.~Vielhauer.
\newblock Artificial steganographic network data generation concept and
  evaluation of de- tection approaches to secure industrial control systems
  against stegano- graphic attacks.
\newblock {\em In The 16th International Conference on Availability, Relia-
  bility and Security (ARES 2021), August 17--20, 2021, Vienna, Austria. ACM,
  New York, NY, USA, 9 pages. https://doi.org/10.1145/3465481.3470073}, 2021.

\bibitem{IFAC2020}
T.~Neubert and C.~Vielhauer.
\newblock Kill chain attack modeling for hidden channel attack scenarios in
  industrial control systems.
\newblock {\em 21st IFAC World Congress, Berlin, Germany, July 11-17,
  Submission 1475}, 2020.

\bibitem{Wendzel_2022}
S.~Wendzel, L.~Caviglione, W.~Mazurczy, A.~Mileva, J.~Dittmann, C.~Kr{\"a}tzer,
  K.~Lamsh{\"o}ft, C.~Vielhauer, L.~Hartmann, J.~Keller, T.~Neubert, and
  S.~Zillien.
\newblock A generic taxonomy for steganography methods.
\newblock July 2022.

\bibitem{ARES21_W}
S.~Wendzel, L.~Caviglione, W.~Mazurczyk, A.~Mileva, J.~Dittmann,
  C.~Kr{\"a}tzer, K.~Lamsh{\"o}ft, C.~Vielhauer, L.~Hartmann, J.~Keller, and
  T.~Neubert.
\newblock A revised taxonomy of steganography embedding patterns.
\newblock {\em The 16th International Conference on Availability, Reliability
  and Security (ARES 2021), August 17--20, 2021, Vienna, Austria. ACM, New
  York, NY, USA, 12 pages. https://doi.org/10. 1145/3465481.3470069}, 2021.

\end{thebibliography}

\end{document}